\documentclass{elsart3p}

\usepackage{graphicx}
\usepackage{amssymb}

\begin{document}

\begin{frontmatter}



\title{A high resolution scintillating fiber tracker with SiPM  readout}

\author{Henning Gast, Thomas Kirn, \textit{Gregorio Roper Yearwood}, Stefan Schael}
\address{Rheinisch-Westf\"alische Technische Hochschule, Aachen, Germany}


\begin{abstract}
A novel modular high-resolution charged-particle tracking detector design using $250\mu m$ diameter, round, scintillating fibers and SiPM arrays for readout is presented. The fiber modules consist of 8 layers of 128 fibers, mechanically stabilized by a carbon fiber / Rohacell foam structure. 
A prototype using scintillating fibers with a diameter of $300\mu m$ and two types of silicon photomultipliers has been tested in a $10GeV$ proton beam in late October 2006 at the T9 PS-beamline, CERN. We present the measured spatial resolution, efficiency and signal-over-noise for this setup. The advantage of this tracker concept is a 
compact and modular design with low material budget and only little or no
cooling requirements depending on the quality of the silicon
photomultiplieres.
\end{abstract}

\end{frontmatter}

% main text
\section{Introduction}
\label{sec::intro}
The idea to use scintillating fibers for charged-particle tracking is not
new. Several experiments have already successfully built and used
scintillating fiber trackers in high-energy and astro-particle physics. The
\textit{D\O\ } experiment \cite{dzero05} for example chose to use a
scintillating fiber tracker built from $835\mu m$ thick, multi-clad
scintillating fibers, read out by \textit{Visible-Light Photon Counters}
(VLPC). The MICE experiment took the idea of a scintillating fiber tracker one
step further, using $350\mu m$ thin multiclad fibers, also read out by
VLPCs. The use of VLPCs requires a significant overhead, because they have to
be operated in cryostats at temperatures of around $7K$. We therefore propose
a new detector design, using $250 \mu m$ thin fibers, read out by
\textit{Silicon Photomultipliers (SiPM)} that can be used very effectively
with only moderate cooling.

\section{Silicon Photomultipliers}

A SiPM is a relatively new kind of multipixel semiconductor photodiode that achieves a high intrinsic gain by being operated above its breakdown voltage\cite{dol01}. SiPMs have a high photodetection efficiency of up to $65\%$, exceeding that of regular photomultiplier tubes. In addition, they scale to small dimensions, allowing for a compact readout of thin scintillating fibers.

Major distributors of SiPM are Hamamatsu\cite{ham07}, Japan and Photonique\cite{photoniqueweb}, Switzerland. Photonique SiPM of type SSPM-0606EXP and SSPM-050701GR were used during the beamtest of a SciFi/SiPM tracker prototype. 

\begin{figure}
\begin{center}
\includegraphics[width=8cm,angle=0]{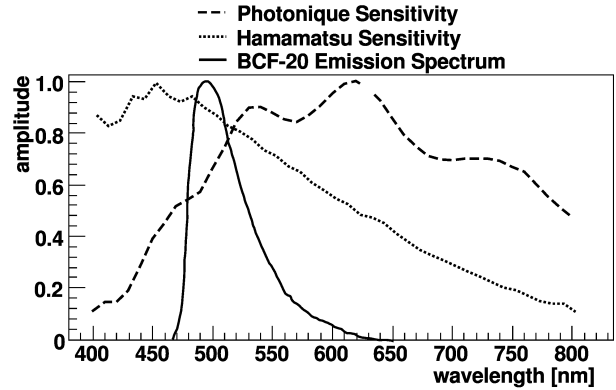}
\end{center}
\caption{Emission spectrum of a BCF-20 fiber \cite{bicron} in comparison to the wavelength dependant response functions of a Hamamatsu SiPM (model MPPC S10361-025U) and a Photonique SiPM (model SSPM-0606EXP) \cite{mus07} }
\end{figure}

SiPM from Photonique reach the peak sensitivity at around $600nm$ (see fig. 1). The type SSPM-050701GR has a very high internal gain greater than $10^6$ while showing a dark count rate of around $1MHz$ at room temperature and about $200kHz$ at $-10^{\circ}C$ at the recommended operating voltage of typically $41V$. Its photodetection efficiency at $494nm$ is $25\%$ (see fig. 2). The type Photonique SSPM-0606EXP, which does not have any protective epoxy layers on top of the device, exhibits a photodetection efficiency of about $40\%$ at a pixel density of 556 pixels per $mm^2$. The gain of the SSPM-0606EXP device is about $3 \cdot 10^5$. The noise of this device is about 5 times higher than that of the SSPM-050701GR device. 

The Hamamatsu model MPPC S10361-025U has an extremely high pixel density of $1600 / mm^2$. The resulting low geometrical fill factor of $37\%$ significantly limits the photodetection efficiency to about $25\%$ at $494nm$. Hamamatsu also offers a model S10361-100U with only 100 pixels per $mm^2$ and a geometrical fill factor close to $80\%$ that offers a photodetection efficiency of $65\%$\cite{ham07}. The dark noise rate of the tested Hamamatsu MPPC S10361-025U is $300kHz$ at $21^{\circ} C$ and around $20kHz$ at $-20^{\circ} C$. 

\begin{figure}
\begin{center}
\includegraphics[width=8cm,angle=0]{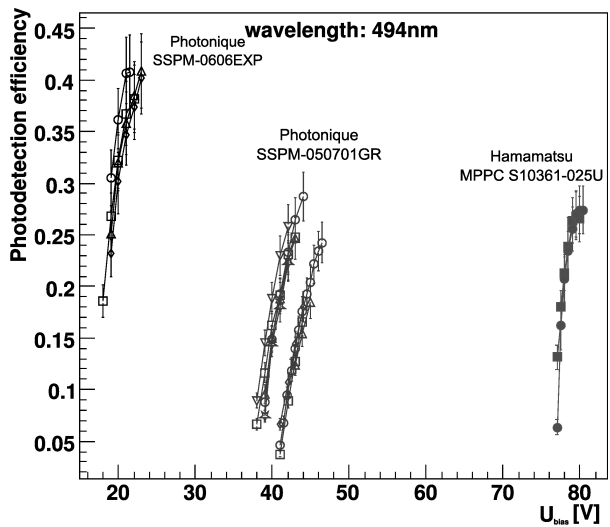}
\end{center}
\caption{The photodetection efficiency of various SiPM types depending on the bias voltage}
\end{figure}

\section{Beamtest of the first Prototype}

\begin{figure}
\begin{center}
\includegraphics[width=8cm,angle=0]{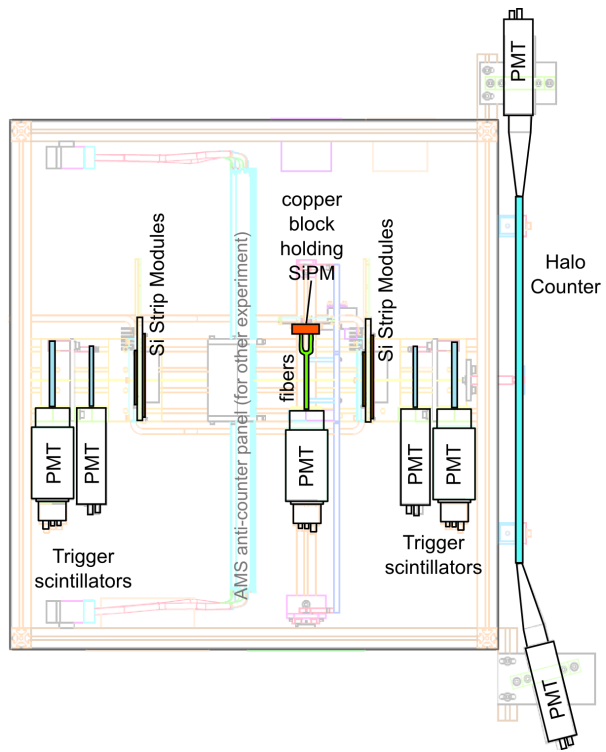}
\end{center}
\caption{A drawing of the beamtest setup shows the triggers, the beam
  telescope and the actual prototype consisting of two fiber ribbons and 19
  SiPM to read out the fibers.}
\end{figure}

The prototype (see fig. 3) consisted of $300\mu m \times 300 \mu m$ square, multiclad fibers of type Bicron\cite{bicron} BCF-20 with white EMA coating and Photonique SiPMs of type SSPM-050701GR and of type SSPM-0606EXP. The peak emission wavelength of BCF-20 fibers is at $492nm$, matching the peak sensitivity of Photonique SSPM-0606EXP SiPM (see fig. 1).

The scintillating fibers were arranged in two ribbons of $3 \times 10$ fibers. The fiber ribbons were stabilized using glue as an adhesive. Both ends of the 3-fiber ribbons were glued into a plastic connector and polished. One end was connected to a SiPM by mounting it into a copper block and held in place by a aluminum frame and a spring.
The SiPM'S were mounted into the copper-block to allow for a temperature control.
Part of the beam test  the opposing end of the fibers was covered by a reflective foil to increase the light output for the SiPMs.

A beam telescope with four silicon strip modules from the CMS tracker project
was used to measure the position of the incident particles. The silicon
modules had a strip pitch of $113..139 \mu m$ and an intrinsic resolution of
about $40 \mu m$ \cite{cms04}.

Four scintillator counters, two in front of and two behind the prototype provided the trigger with a halo counter as a veto.

\begin{figure}
\begin{center}
\includegraphics[width=8cm,angle=0]{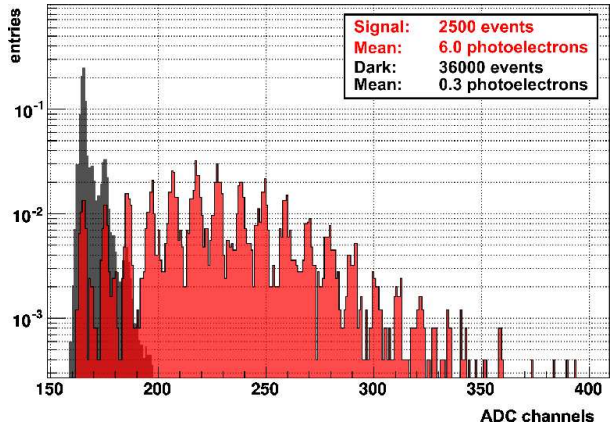}
\end{center}
\caption{The signal spectrum and the dark spectrum of a SSPM-0606EXP device during the beamtest with reflective foil. Both spectra are normalized to 1.}
\end{figure}

The beamtest of the prototype took place in a 10GeV proton beam at PS,
CERN. During the beamtest, 1.5 million events were recorded and about 800,000
particle tracks were reconstructed with the beam telescope. The position of
each fiber column was determined by reconstructing the position of particles
that produced a high signal within the fiber (see fig. 4). The average
measured distance between two fiber columns was $309\mu m$ with a precision of
$10\mu m$. The spatial resolution for particles of perpendicular incidence
that pass through all three fibers of one fiber column was about $90\mu m$
which matches the expected intrinsic resolution of $\frac{d}{\sqrt{12}}$ where
d is the fiber pitch.

\begin{figure}
\begin{center}
\includegraphics[width=8cm,angle=0]{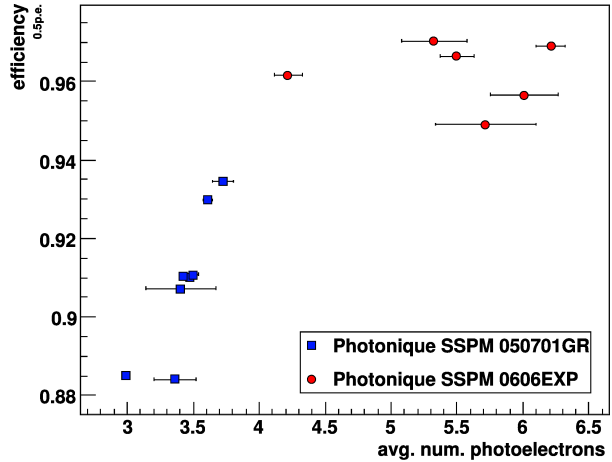}
\end{center}
\caption{A plot of the measured efficiency of the particle detection vs. the measured average number of photoelectrons for new and old Photonique SiPM. The efficiency was determined for the lowest possible cut of $0.5 p.e.$ for the SiPM signal.}
\end{figure}

Knowing the positions of the fibers, we determined the average photoelectron yield for particles that passed through one of the fiber columns. For particles with perpendicular incidence the average photoelectron yield for both types of SiPM with and without reflective foil on the opposing fiber end was measured. The SSPM-050701GR signal was about $2.2$ photoelectrons without and $3.5$ photoelectrons with reflective foil (see fig. 5). The SSPM-0606EXP achieved an average photoelectron yield of $3.8$ without reflective foil and $5.8$ with reflective foil (omitting one SiPM that actually showed a reduced photoelectron yield after adding the reflective foil).

The measured signal-to-noise ratio was about 20 for the SSPM-0606EXP with reflective foil and about 100 for the SSPM-050701GR with reflective foil. The mean efficiency for perpendicular incidence, setting a cut at 0.5 photoelectrons, was $96\%$ for the ribbon read out by the SSPM 0606EXP and $91\%$ for the SSPM 050701GR. 

\section{Tracker Module Design}
\begin{figure}
\begin{center}
\includegraphics[width=8cm,angle=0]{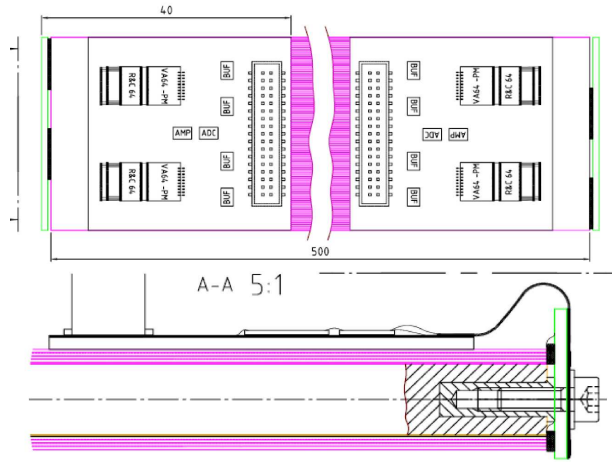}
\end{center}
\caption{Drawing of the geometry of a single tracker module, showing the fiber
  placement on the mechanical support structure and the frontend hybrids
  mounted on the fiber module.}
\end{figure}

The tracker design is modular. It consists of several layers of tracker modules, each module consisting of 8 layers of scintillating fibers with 128 fibers in every layer. 4 layers of fibers are glued to each side of a module core that consists of $5mm$ thick Rohacell foam covered by $100\mu m$ thin carbon fiber skins on either side of the module (see fig. 6). Neighboring layers are shifted by one half of the fiber pitch with respect to each other to improve the spatial resolution. 

SiPM arrays with a sensitive area of $8mm \times 1mm$ and 32 readout channels, each channel covering an area of $0.25mm \times 1mm$ are used for column-wise fiber readout. The SiPM arrays are mounted on alternating ends of the fiber modules along with an integrated preamplifier and digitization solution. The opposing ends of the fibers are covered by a reflective coating. 

\begin{figure}
\begin{center}
\includegraphics[width=8cm,angle=0]{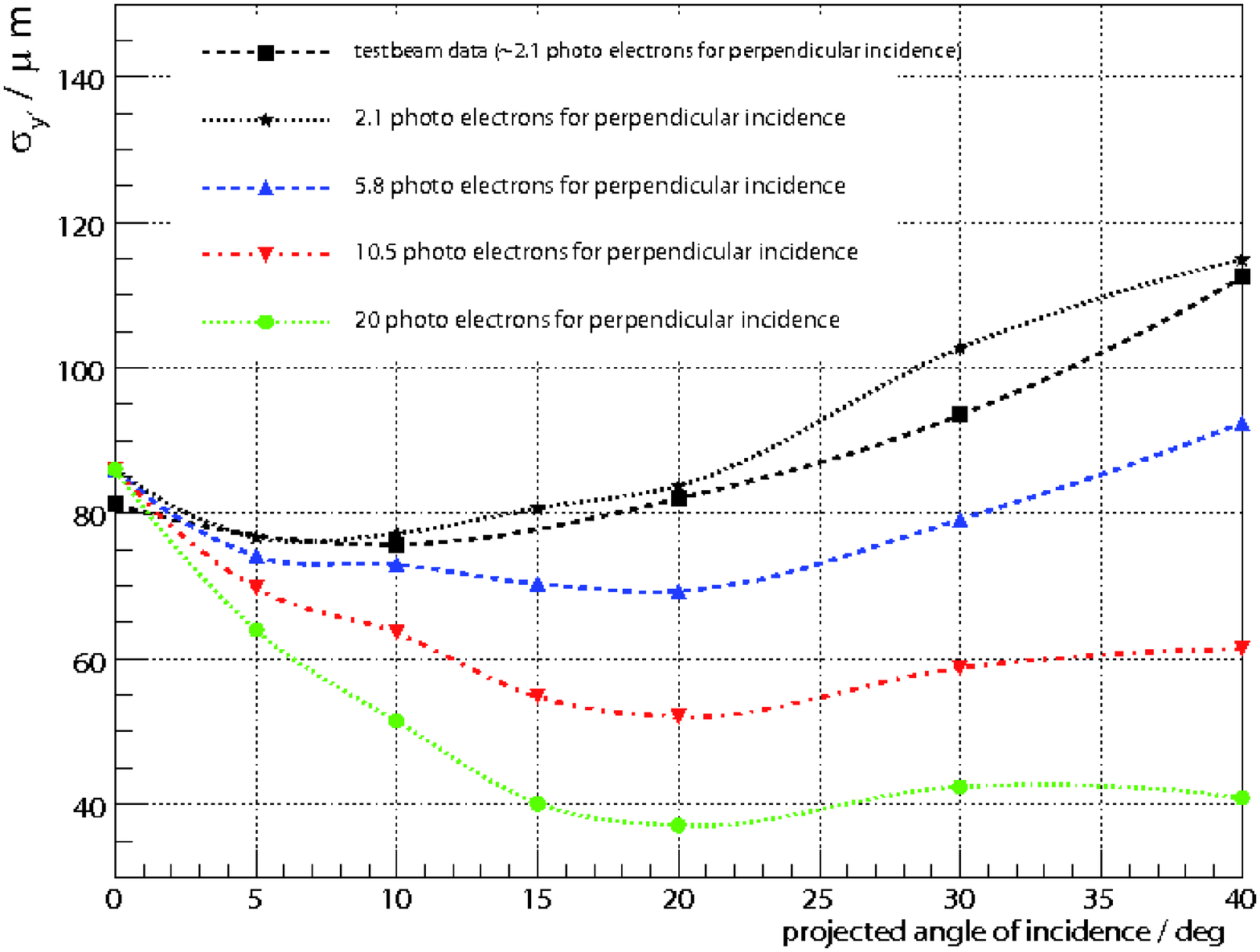}
\end{center}
\caption{Spatial resolution for a bundle of fibers of $300\,\mu{}m$
width from testbeam data and Monte Carlo simulations. Testbeam data
obtained with a fiber bundle without reflective foil and Photonique
SSPM-050701GR SiPM
readout are plotted using square markers. Results from
Monte Carlo simulations are added to study the behavior for improved
photo electron yields. A yield of $5.8$ photo electrons was reached in
the testbeam with SSPM-0606EXP SiPMs and reflective foil, but only
data at $0^\circ$ were taken in this configuration.}
\end{figure}
A dedicated Monte Carlo simulation, again using the Geant4 package,
has been developed for comparison to and generalization of the
testbeam results. A key question to be answered was the spatial
resolution obtained with a fiber module as a function of the mean photo
electron yield $n_{p.e.}$ of the fibre-SiPM chain. 
Figure~7 shows the result. The spatial resolution $\sigma_{y^\prime}$ is plotted
for different values of $n_{p.e.}$ and depending on the angle $\alpha$ of incidence
of a particle, projected into the bending plane of the
magnet. $\sigma_{y^\prime}$ is the resolution along the axis
perpendicular to the fibers. Since the beam telescope used in the
testbeam measured the coordinate $y$ perpendicular the direction of
incidence $z$, $\sigma_{y^\prime}$ is calculated from the measured
$\sigma_y$ and the positioning accuracy $\sigma_z=10\,\mu{}m$ as
follows:
\begin{equation}
\sigma_{y^\prime}=\sigma_y \cos\alpha\oplus
\sigma_z\sin\alpha
\end{equation}
For the photo electron yield achieved in the testbeam, a spatial
resolution of $72\,\mu{}m$ is obtained at the mean projected angle of
incidence, which is $\bar{\alpha}=11^\circ$ for the PEBS geometry.

\section{Conclusion}
The testbeam results indicate that this concept for a high-resolution SciFi/SiPM tracker is technical feasible. The average yield of 6 photoelectrons 
with a reflective foil on one fiber end and the SSPM-0606EXP SiPM matched our requirements. SiPMs with a reduced pixel density and a $50\%$ higher PDE are already 
commercially available from Hamamatsu. 
Furthermore the light output from scintillating fibers can be improved by $20\%-40\%$ using fibers without white coating as measurements with different fiber coatings conducted for the CREAM experiment have shown\cite{cre05}. Up to now we have not engineered the optical coupling
between fibers and SiPM's at all.

A spatial resolution as good as $40 \mu m$ is in principle possible, depending on the
granularity of the readout, the quality of the SiPM's, the qualtity of the optical coupling of the 
fibers to the SiPM's and the type of fibers used. With the understanding and technology we have today,  an average spatial resolution of $60 \mu m$ is expected.


\begin{thebibliography}{00}


% notes:

% \bibitem{label2}
% If there is a note, it should come last:
% \bibitem{label3} \note
% \end{subbibitems}
\bibitem{dzero05} D\O\ Collaboration, NIM A565, p. 463-537, 2006
\bibitem{yos06} Yoshida, M., \textit{MICE Overview - Physics Goals And Prospects}, Proceedings of EPAC 2006, Edinburgh, Scotland
\bibitem{dol01} Dolgoshein, B. et al., NIM A504, p. 48-52, 2003 
\bibitem{ham07} Hamamatsu Photonics, K.K., Japan URL: http://sales.hamamatsu.com
\bibitem{photoniqueweb} Photonique SA, Switzerland URL: http://www.photonique.ch
\bibitem{bicron} Saint-Gobain Crystals, Paris, France URL: http://www.bicron.com
\bibitem{cms04} The CMS Collaboration, NIM A517, p. 77–93, 2004
\bibitem{mus07} Musienko, Y., Unpublished Measurements for the PEBS project at APDlab, CERN, Switzerland, 2007
\bibitem{cre05} Young Soo Yoon et al., 29th ICRC Pune, p. 101-104, 2005



\end{thebibliography}
\end{document}